\documentstyle[sprocl]{article}

\def\be{\begin{equation}}
\def\ee{\end{equation}}
\def\bea{\begin{eqnarray}}
\def\eea{\end{eqnarray}}

\def\Lam{\Lambda}
\def\Sig{\Sigma}
\def\DI{\Delta I}
\def\pip{\pi^+}
\def\pim{\pi^-}
\def\piz{\pi^0}
\def\vs{{\it v.s.}}
\def\ie{{\it i.e.}}

\begin{document}

\title{CHIRAL SYMMETRY AND WEAK DECAY OF HYPERNUCLEI}

\author{Makoto OKA}

\address{Dept. of Physics, Tokyo Institute of Technology\\ 
Meguro, Tokyo 152-8551, Japan\\E-mail: oka@th.phys.titech.ac.jp}

\maketitle\abstracts{The weak decays of hyperons and hypernuclei are
studied from the chiral symmetry viewpoint.  The soft pion relations
are useful in understanding the isospin properties of the weak hyperon
decays.  Recent development on the short-range part of the $\Lam N\to 
NN$ weak transitions shows fairly good account of the weak decays
of hypernuclei, though it fails to explain the $n/p$ ratio.
The $\pip$ decays of light hypernuclei are studied in the soft pion
approach.  They are related to the $\DI=3/2$ amplitudes of the 
nonmesonic decay.}

\section{Introduction}
\def\HPV{H^{PV}}
\def\HPC{H^{PC}}
\def\fpi{f_{\pi}}

While chiral symmetry is a powerful tool in understanding properties 
of low lying hadrons and their interactions, its role in the weak 
decays of hypernuclei has not been explored as far as we know.  Here 
we suggest that chiral symmetry is significant in understanding weak 
mesonic decays of hyperons in the free space and also in nuclear 
medium.  On the other hand, it was recently shown that the short 
range part of the weak $\Lam N\to NN$ transition is attributed to the 
direct quark  processes, and numerical calculation suggest that it is
significant in nonmesonic weak decay of hypernuclei.{\cite{ITO,IOMI}}

\section{Chiral Symmetry}
What is the role of chiral symmetry in the hyperon decay?
A useful tool to take account of the chiral symmeric dynamics of 
hadrons is the soft pion theorem,{\cite{CA}}
\be
\langle \alpha \pi^a(q) | \hat O |\beta \rangle 
\stackrel{q^{\mu}\to 0}{\longrightarrow}
-\,{i\over \fpi} \langle \alpha|[Q^{a}, \hat O] |\beta\rangle 
+ (\hbox{pole terms})
\label{eq:soft-pi}
\ee
This can be applied to the weak pionic processes,such as,
\be
\langle n\piz(q) | H^{PV} |\Lam \rangle \stackrel{q^{\mu}\to 0}{\longrightarrow}
-\,{i\over \fpi} \langle n|[Q^0_{5}, \HPV] |\Lam\rangle =
-\,{i\over 2\fpi} \langle n| \HPC |\Lam\rangle
\label{eq:lam-decay}
\ee
Here $\HPV$ is the parity violating part of the weak hamiltonian, 
which contains only the left-handed currents $\bar q_{L}^a\gamma^{\mu} 
q_{L}^a$ and the flavor singlet right handed current induced by the 
penguin type QCD corrections.  This allows us to relate the
commutation $[Q^0_{5}, \HPV]$ in the second expression of 
eq.(\ref{eq:lam-decay}) to $\HPC$ in the final expression as
\bea
 \,[Q^a_{R}, H_{W}] &=& 0 \nonumber\\
 \,[Q^a_{5}, H_{W}] &=& -[Q^a_{L}, H_{W}] = -[I^a, H_{W}] \nonumber \\
\langle n|[Q^a_{5}, \HPV]|\Lam\rangle &=& \langle n|[I^a, \HPC] |\Lam\rangle
= -{1\over 2} \langle n| \HPC |\Lam\rangle  \nonumber
\eea
Thus, the parity violating amplitudes, or the $S$-wave decay amplitudes, 
of pionic decays of various hyperons can be expressed in terms of the
baryonic matrix elements of the parity conserving weak hamiltonian.
As a result, for instance, the $\DI=1/2$ dominance of $\Lam\to N\pi$ 
decays follows immediately since ``$\Lam\to n$ transition'' is purely 
$\DI=1/2$.  Furthermore if we conjecture that $H_{W}$ is purely flavor 
octet, then various matrix elements, 
$\langle n| \HPC |\Sig\rangle$, $\langle \Sig| \HPC |\Xi\rangle$,
are all related to $\langle n| \HPC |\Lam\rangle$ and thus several 
relations among the pionic decay amplitudes of hyperons are obtained.  
Such relations are known to be satisfied fairly well for the $S$-wave 
decay amplitudes of the hyperon decays.{\cite{Dono}}

The parity conserving decays belong to exceptions of the soft pion 
theorem, in which the pole terms cannot be neglected. 
The pole terms are such that $\Lam\to n \to n\piz$ or
$\Lam\to \Sig^0\piz \to n\piz$, and their amplitudes are
\bea
\langle n\piz(q\to 0)| \HPC |\Lam\rangle &\sim& \langle n\piz|n\rangle 
{i\over m_{\Lam}-m_{n}} \langle n| \HPC |\Lam\rangle  \nonumber\\
&+& \langle n| \HPC |\Sig^0\rangle {i\over m_{n}-m_{\Sig}}
\langle \Sig^0\piz |\Lam\rangle
\eea
Assuming the pole dominance of the parity conserving, $P$-wave, decay 
amplitudes, we find interesting relations of the $\Sig^+\to n\pip$ decay.
It can be easily shown that the soft-pion amplitude vanishes,
\be
\langle n\pip| \HPV |\Sig^+ \rangle_{\hbox{soft-pion}} = 0
\ee
since 
\be
[I^-, \HPC] =0 \nonumber
\ee
for the $\DI=1/2$ dominant $\HPC$, which has $\DI_{3}=-1/2$.
On the other hand, there is no such constraint for the pole terms of 
the parity conserving amplitudes, where the $p$, $\Sig^0$ and 
$\Lam$ intermediate states with different energy denominators 
contribute.
Therefore the soft-pion theorem suggests that the $\Sig^+\to n\pip$ 
decay goes only 
through the parity conserving $P$-wave channel.
This is indeed what we observe experimentally,  the PV amplitude 0.13 
\vs\ the PC 42.2.
This example shows that the $\DI=1/2$ dominance and the soft-pion relation
is very well satisfied in this decay.

We later consider $\pip$ decays of hypernuclei in this context and see
that the soft-pion theorem suggests the $\pip$ decays are induced 
only by the $\DI=3/2$ part of the weak hamiltonian.

\section{$\DI=1/2$ Rule}
\def\ubar{\bar u}
\def\dbar{\bar d}
\def\qbar{\bar q}

In the above discussion, we have assumed that $\DI=1/2$ dominance of 
the weak matrix element, $\langle N| \HPC |\Sig\rangle$.
Explanation of the ``$\DI=1/2$ rule'' has been a long standing problem 
of the weak decay of the kaons and the hyperons.  
It was shown long time ago that the perturbative QCD corrections to 
the standard model weak vertex enhance the $\DI=1/2$ component, while 
it suppresses the counterpart, $\DI=3/2$ component.
The mechanism can be understood easily by decomposing the weak 
$s+d\to u+d$ transition into the isospin-spin-color eigenstates.
As the strangeness changing transition is induced only by the charged 
current, or the $W$-boson exchange, the vertex at low energy is given 
without the QCD corrections by
\be
(\ubar^{\alpha}_{L}\gamma^{\mu}s^{\alpha}_{L})
(\dbar^{\beta}_{L}\gamma_{\mu}u^{\beta}_{L})
= (\dbar^{\beta}_{L}\gamma^{\mu}s^{\alpha}_{L})
  (\ubar^{\alpha}_{L}\gamma_{\mu}u^{\beta}_{L})
\ee
where $\alpha$ and $\beta$ are color indices and
the equality comes from the Fierz transformation.
From this we observe that the color+isospin combination of the final 
$u+d$ quarks is always symmetric, namely, 
($I_{f}=0$, Color ${\bf\bar 3}$) or
($I_{f}=1$, Color ${\bf 6}$).
In both cases, the total spin of the final quarks must be 0.
When we consider gluon corrections to this vertex, we notice that the 
gluon exchange, or its color-magnetic component, 
$- (\lambda_{1}\cdot\lambda_{2})(\sigma_{1}\cdot\sigma_{2})$ term,
between the final $u$ and $d$ is attractive for
($I_{f}=0$, Color ${\bf\bar 3}$), 
\be
\langle S=0, C={\bf\bar 3}|(\lambda_{1}\cdot\lambda_{2})
(\sigma_{1}\cdot\sigma_{2})
|S=0, C={\bf\bar 3}\rangle= -8  ,
\ee
while it is repulsive in the other,
\be
\langle S=0, C={\bf 6}|(\lambda_{1}\cdot\lambda_{2})
(\sigma_{1}\cdot\sigma_{2})
|S=0, C={\bf 6}\rangle= +4  .
\ee
Therefore the QCD correction tends to enhance the final $I_{f}=0$ amplitude.
The above heuristic explanation of the $\DI=1/2$ enhancement can be 
confirmed in the renormalization group improved effective action of 
the strangeness-changing weak interaction.{\cite{hweak}}  
It was also shown that a
further enhancement of $\DI=1/2$ is resulted due to the ``Penguin'' 
diagrams, which is regarded as a QCD-corrected $s\to d$ transition, 
and is purely $\DI=1/2$.{\cite{hweak}} 

The perturbative QCD corrections are not the only source of the 
enhancement, but we expect further effects due to nonperturbative 
origin.  In fact, it is known that the enhancement in the effective 
interaction is not large enough to explain the observed dominance of 
$\DI=1/2$.  We here concentrate on the baryonic weak interaction.
Miura-Minamikawa and Pati-Woo{\cite{MMPW}} pointed out that the 
$\DI=3/2$ part
of the nonleptonic hyperon decays is suppressed due to the color 
symmetry of the valence quarks in the baryon.  It is understood 
easily by considering the color structure of the 4-quark operator 
which belongs to the 27-dimensional irrep.\ of the flavor $SU(3)$:
\be
O({\rm 27}) = (\ubar^{\alpha}_{L}\gamma^{\mu}s^{\alpha}_{L})
(\dbar^{\beta}_{L}\gamma_{\mu}u^{\beta}_{L} )
+ (\dbar^{\alpha}_{L}\gamma^{\mu}s^{\alpha}_{L})
  (\ubar^{\beta}_{L}\gamma_{\mu}u^{\beta}_{L})
  \label{eq:O27}
\ee
which is the part responsible for the $\DI=3/2$ transition.
From the symmetry of the final $u$ and $d$ quarks, this operator 
creates two quarks with their color part being symmetric, \ie, the color 
${\bf 6}$ state. As the color wave function for the valence three 
quarks of the baryons, either the hyperon in the initial state or the 
nucleon in the final state, this operator cannot be connected to two 
quarks inside the baryon.  Thus, the only possibility is the external 
diagram in which both the quark and the antiquark of the meson (pion)
are connected to the weak vertex directly.  
Such diagram is not allowed either because the external $q\qbar$ 
created by eq.(\ref{eq:O27}) should be both left-handed and therefore 
does not make a pseudoscalar meson in the chiral limit.
Gluon exchanges among the initial or final quarks do not help,
while ``exotic'' component such as valence gluon will change the 
situation.
Thus we observe that the $\DI=3/2$ part of the $Y\to N+ \hbox{PS meson}$ 
is strongly suppressed as far as we consider the valence quark picture 
of the baryons. 
It should be noted, however, that the vector 
mesons may couple with $\DI=3/2$ verteces
directly.{\cite{MS2}}

Another possibility is the $\DI=1/2$ enhancement due to the diquark 
components in the baryon. It is an enhancement of  $0^+$
$ud(I=0)$ diquark components, which is favored by the gluon exchanges.
This is in fact the same mechanism as the $\DI=1/2$ enhancement in the 
perturbative correction discussed above.  We, however, have found that 
such effect may be small because the naive evaluation of the $\HPC$ 
in the harmonic oscillator valence quark model gives strong enough 
transition amplitude  for the $\Lambda\to N\pi$ decay according to the 
soft-pion formula, eq.(\ref{eq:lam-decay}).

\section{Direct Quark Mechanism and Weak Decay of Hypernuclei}

The pionic decay of $\Lam$ is known to be suppressed in nuclear 
medium as the final nucleon do not have enough momentum to go above 
the Fermi energy.
Thus the main decay mode is nonmesonic, which can be described by the 
simplest elementary processes,
$\Lam p \to pn$ and $\Lam n \to nn$.
These processes are viewed as weak baryonic interactions, which is 
unique and interesting itself as a 
new type of the nonleptonic weak interactions of baryons.
Furthermore, this is a reaction in which the momentum transfer is so 
large that the quark substructures of the baryons may be significant.

Recently, we proposed the direct quark (DQ) transition mechanism to 
account for the short-range part of the $YN\to NN$ weak 
interaction.{\cite{ITO,MS}}
The DQ transition potential is obtained by evaluating $su\to ud$ and 
$sd\to dd$ transitions among the valence quarks in two baryons.

The decay rates of the $\Lambda$ in nuclear matter, and in light 
hypernuclei are calculated with the transition potential, which 
includes DQ and $\pi$ and $K$ meson exchanges.  The 
results are compared with those without DQ, and also with experiment.
We leave the details to literature,{\cite{Sasaki}} 
while the conclusions of our study 
are summarized here.
(1) The DQ transition is significantly large, and shows qualitative 
differences from the meson exchanges.
(2) The one pion exchange (OPE) mechanism yields a large tensor amplitude, 
\ie, 
the transition from $\Lambda p : ^3S_{1}$ to $np: ^3D_{1}$.  
This seems in fact too 
large if we employ a hard form factor as is used in the Bonn potential.
We have pointed out that softer form factors for the pion-nucleon 
couplings are more appropriate.  
(3) The large tensor amplitude in OPE causes a difficulty that 
the $nn/np$ ratio is too small compared to the value suggested from
experiment, so-called the $n/p$ ratio problem.  
We found that this problem is not solved completely only by the 
introduction of a soft form factor for OPE, but it is clear that 
the short-range mechanisms of the weak transition are important.
Indeed, the DQ amplitude enhances the $nn$ decay rate significantly 
and therefore improves the ratio.
(4) The $\DI=3/2$ contribution is significant for the $J=0$ transition 
amplitudes in the DQ mechanism, while the meson exchanges are assumed 
purely in the $\DI=1/2$ transition.  It should be noted here that the 
mechanisms to enhance the $\DI=1/2$ amplitudes in other hadronic weak 
interactions are not effective here and therefore the $\DI=3/2$ 
amplitudes may be comparable to the $\DI=1/2$ one.
Unfortunately, it is not possible at present to determine the 
importance of the $\DI=3/2$ contribution from the experimental 
data.{\cite{Schu}}

\section{$\pi^{+}$ decay mode and $\DI=3/2$ amplitudes}

Light hypernuclei may decay weakly by emitting a pion.  While the free $\Lam$ 
decays into $p\pim$ or $n\piz$, the $\pip$ decay requires an assistance
of a proton, \ie, $\Lam+p\to n+n+\pip$.
Some old experimental data suggest that the ratio of $\pip$ and 
$\pim$ emission from $^4_{\Lambda}He$ is about 5\%.{\cite{pip-data}}
The most natural explanation of this process is $\Lam\to n\pi^0$ decay 
followed by
$\pi^0 p \to \pip n$ charge exchange reaction.
It was evaluated for realistic hypernuclear wave functions and
found to explain only 1.2\% for the $\pip/\pim$ ratio.{\cite{DHCG}}
Another possibility is to consider $\Sig^+ \to \pip n$ decay
after the conversion $\Lam p \to \Sig^+ n$ by the strong interaction
such as pion or kaon exchanges.
It was found, however, that the free $\Sig^+$ decay which is dominated
by $P$-wave amplitude, gives at most 0.2\% for the $\pip/\pim$ ratio.

\let\di=\DI

In order to solve this problem, we have applied 
the soft-pion technique to the $\pip$ decay of light
hypernuclei.{\cite{piplus}}
The soft-pion theorem to the process $\Lam p \to nn \pip(q\to 0)$
reads
\begin{equation}
  \lim_{q\to 0} \langle nn\pip(q)|H_W|\Lam p\rangle 
= -{i\over \sqrt{2} f_{\pi}} \langle nn|[Q_5^-, H_W]|\Lam p \rangle
\label{soft-pion}
\end{equation}
Again, because of
\begin{equation}
  [Q_5^-, H_W] = -[ I_-, H_W]
\end{equation}
it discriminates the isospin properties of $H_W$.
Similarly to the case of $\Sig^{+}$ decay, we see
that the $\di=1/2$ part vanishes as
\begin{eqnarray}
  [I_-,H_W(\di=1/2, \di_z=-1/2)] &=& 0 \\{}
  [I_-,H_W(\di=3/2, \di_z=-1/2)] &=& \sqrt{3} H_W(\di=3/2, 
  \di_3=-3/2) 
\end{eqnarray}
We then obtain
\begin{equation}
  \lim_{q\to 0} \langle nn\pip(q)|H_W|\Lam p\rangle 
= {i\sqrt{3} \over \sqrt{2} f_{\pi}} 
   \langle nn|H_W(\di=3/2, \di_3=-3/2)|\Lam p \rangle
   \label{eq:pipamp}
\end{equation}
Thus we conclude that the soft $\pip$ emission in the 
$\Lam$ decay in hypernuclei is caused only by the $\di=3/2$ 
component of the strangeness changing weak hamiltonian.
In other words, the $\pip$ emission from hypernuclei probes the 
$\di=3/2$ transition of $\Lam N \to NN$.



\section{Conclusion}

In this article, we have tried to demonstrate how the soft-pion 
approach is useful in understanding the weak hyperon transitions.  
It is amazing that the chiral symmetry plays so important role even 
in the weak processes.  We, however, have further remaining problems.  
It is necessary to go beyond the chiral limit, so that the finite pion 
mass effects are to be included.  Corrections due to the flavor 
$SU(3)$ breaking may also be important.  These can be included by 
the chiral perturbation theory approach.  In view of the qualitative 
success in the soft-pion approach, it is promising to apply the chiral 
effective theories to the hyperon and hypernuclear decay processes.
Several recent papers attempt such approaches for the hyperon decays
with some success.{\cite{chPT}}
It is interesting to study the hypernuclear decays 
in such formulations.

\def \vol(#1,#2,#3){{{\bf {#1}} (19{#2}) {#3}}}
\def \NP(#1,#2,#3){Nucl.\ Phys.\          \vol(#1,#2,#3)}
\def \PL(#1,#2,#3){Phys.\ Lett.\          \vol(#1,#2,#3)}
\def \PRL(#1,#2,#3){Phys.\ Rev.\ Lett.\   \vol(#1,#2,#3)}
\def \PRp(#1,#2,#3){Phys.\ Rep.\          \vol(#1,#2,#3)}
\def \PR(#1,#2,#3){Phys.\ Rev.\           \vol(#1,#2,#3)}
\def \PTP(#1,#2,#3){Prog.\ Theor.\ Phys.\ \vol(#1,#2,#3)}
\def \ibid(#1,#2,#3){{\it ibid.}\         \vol(#1,#2,#3)}
\def\MO{M.~Oka} \def\etal{{\it et al.}}

\end{document}